\documentclass[final,5p,times,twocolumn]{elsarticle}
\usepackage{amssymb}
\usepackage{lineno}
\usepackage[utf8]{inputenc}
\usepackage[T1]{fontenc}
\usepackage{epsfig}
\usepackage{amsmath}
\usepackage{amsfonts}
\usepackage{amssymb}
\usepackage{color}
\usepackage{graphicx}
\usepackage{url,hyperref}
\usepackage{latexsym}
\usepackage{longtable}
\usepackage{float}
\usepackage{tikz}
\usepackage{epstopdf}
\usepackage{url}
\usepackage{caption}

\begin{document}

\begin{frontmatter}
\title{{\mbox{\small
{}}}\textbf{\  Numerical Thermodynamics of  AdS Black Holes with Higher-Order Mass Corrections}}
\author{ Saad Eddine Baddis$^{1}$}
\author{ Adil Belhaj$^{1}$}
\author{ Hajar Belmahi$^{1,2}$}
\cortext[mycorrespondingauthor] {Corresponding author}
\ead{belmahi.hajar@ucd.ac.ma }
\author{Yassine Hassouni$^{1}$}
\author{Maryem Jemri$^{1}$}
\ead{maryem.jemri@um5r.ac.ma }
\address{
$^{1}$ {ESMaR, Faculty of Science, Mohammed V University in Rabat, Rabat, Morocco.}\\
$^{2}$ {National School of Applied Sciences (ENSA), Chouaib Doukkali University, El Jadida, Morocco.} }

\begin{abstract}

Inspired by noncommutative geometry and quintessence, we construct a new class of asymptotically anti-de Sitter black hole solutions characterized by a higher-order mass deformation term $\frac{aM^{n-1}}{r^n}$ involving two parameters, $a$ and $n$. Due to analytical difficulties, we approach the underlying thermodynamics through parallel programming simulations using CUDA codes. By considering models with $n=3$,  $n=4$, and $n=5$, we numerically study certain thermodynamic properties, including the  stability,  the $P$-$v$ criticality, and  the phase transitions. 
\end{abstract}

\begin{keyword}
AdS black holes, Thermodynamics,   Mass correction scenarios,  Parallel programing simulations.
\end{keyword}
\end{frontmatter}

\section{Introduction}
The search for an understanding of the universe has been profoundly marked by significant discoveries, in which mathematical  developments  have progressively become physical realities.  In this regard, black holes are among the most fascinating examples \cite{1,2,3}. Initially seen as an unusual consequence of the mathematical description of  the spacetime, they have gradually become a relevant piece  of modern astrophysics and high-energy physics.  Such  fascinating objects offer a theoretical laboratory for exploring the interplay between many  branches in  physics including  gravity, thermodynamics, quantum  field theory, quantum information,  and condensed  matter issues \cite{4,5,6,7,07}. 

The rich and multifaceted nature of black holes, combined with their connections to such branches of physics, have given rise to a wide range of empirical and  rational   investigations. Among these, the thermodynamics of black holes has attracted considerably interest \cite{70}.  Considering black holes as thermodynamic systems endowed with   certain relevant quantities   and obeying laws analogous to those of classical thermodynamics reveals a remarkably rich phenomenology by identifying  the cosmological constant with  the pressure.   Especially,  promising results have been  found  in the context of anti-de Sitter (AdS) black holes exhibiting  a variety of phase   structures, including  the Hawking-Page transition and  the phase transitions between small and large black holes\cite{8,9,10}. Additionally, the existence of  the critical points, the  universal critical behavior, and the associated scaling laws have been extensively investigated  providing  a better understanding of  black hole thermodynamic issues\cite{11,110,12,13}.

Recently, a  considerable attention has also been devoted to black hole solutions emerging from modified spacetime, particularly  noncommutative geometry (NC) and $\kappa$-deformed spacetime with Lie algebraic structure aspects  \cite{14,014,0114,15,015}.   In NC geometry context, the  implementation  of a fundamental noncommutativity  scale modifies the black hole solutions and consequently affects their thermodynamic quantities, including the Hawking temperature, the  entropy, and the  heat capacity.  Such   mass corrections  have motivated investigations into their thermodynamic stability and possible phase transitions. Similarly,  the $k$-deformed black holes have been explored from a thermodynamic perspective, with particular emphasis on the effect of the mass  deformation parameter on their temperature, entropy, heat capacity, and thermodynamic behaviors \cite{15}. Such  activities  have offered an interesting setting  for understanding how modifications of the underlying spacetime structure influence  the black hole  thermodynamics and its associated phase structures.

Motivated by NC geometry and quintessence,  we construct a new class of asymptotically AdS black hole solutions featuring  higher-order mass deformations $\frac{aM^{n-1}}{r^n}$  carrying  two  parameters $a$ and $n$.  Due to analytical issues,   we approach the underlying thermodynamics   via parallel programing simulations with CUDA codes.   Considering models  with  $n=3$, $n=4$, and $n=5$, we numerically investigate  certain   thermodynamic properties including the   stability, the  $P$-$v$ criticality and the phase transitions.

The organization of this paper is follows.  Section 2  exposes a  static, spherically symmetric solutions by solving the Einstein field equations in the presence of appropriate matter sources. In Section 3, we  investigate their thermodynamic properties  using numerical methods based on parallel programing simulations  with the help of  CUDA codes.  Section 4 explores the phase structure and  the critical behavior of three distinct models. The last section  provides  concluding remarks.

\section{The model}
In this section, we would like to propose a new solution of AdS black holes with a higher-order mass  deformation term. Inspired and motivated by NC  geometry and quintessence, we  provide   an extra term in the metric function of a static and spherically symmetric space-time by solving the Einstein field equations with matter contributions. To start, we consider the line element used in the building of  the deformed black hole geometry  
\begin{equation}
ds^2=
-f(r)dt^2+\frac{1}{f(r)}dr^2+r^2 d\theta^2+r^2\sin^2\theta \, d\phi^2.
\end{equation}
where $f(r)$ is a radial function. 
To determine  such a function   carrying the deformation mass term that we are after, we need to  handle the  Einstein equations of motion  with matter contributions 
\begin{equation}
G^\mu_\nu + \Lambda \delta^\mu_\nu= 8\pi T^\mu_\nu,
\end{equation}
where $G^\mu_\nu$ is the Einstein tensor and $\Lambda$ is the cosmological constant. $T^\mu_\nu$ represents
the energy-momentum tensor given by
\begin{equation}
T^\mu_\nu=diag(\rho,p_r,p_t,p_t), 
\end{equation}
witch describes  different deformation scenarios.  Inspired by  quintessence \cite{16}, we propose the following ansatz
\begin{align}
\rho=& \dfrac{ n\, \alpha}{8\, \pi \,r^{n+3}}\\
p_r=& -\dfrac{ n\, \alpha}{8\, \pi\, r^{n+3}}\\
p_t=& \dfrac{ n\, (n+1)\,   n\, \alpha}{16\, \pi\, r^{n+3}},
\end{align}
where $n$ is a dimensionless positive parameter. $\alpha$ is a physical parameter with the following dimension
\begin{equation}
\left[ \alpha\right] = \left[ r\right]^{n+1}=L^{n+1} 
\end{equation}
where $L$ denotes the length dimension. A close examination shows that $\alpha$ can be considered as a coupling quantity involving black hole internal parameters. A priori in black hole physics, there are many coupling scenarios by implementing either the mass or the charge of the black hole. Since we are considering a neutral solution, we could exploit the black hole mass term. Inspired by recent results on NC black holes, we offer the following coupling  mass term 
 \begin{equation}
 \alpha=aM^n,
\end{equation}
where $a$ is a relevant modification parameter with the  length dimension
 \begin{equation}
 \left[ a \right] = \left[ r \right]=L,
\end{equation}      
encoding the spacetime geometric deformations.  In this way,  $M$ can be identified with the mass of the point like source of  the gravitational field.  In this mass coupling scenario, the above ansatz becomes
\begin{align}
\rho=& \dfrac{a\, n\, M^n}{8\, \pi \,r^{n+3}}\\
p_r=& -\dfrac{a \, n\, M^n}{8\, \pi\, r^{n+3}}\\
p_t=& \dfrac{a\, n\, (n+1)\,  M^n}{16\, \pi\, r^{n+3}}.
\end{align}
 Putting these forms in the Einstein field equation, we get the following equations of motion
\begin{align}
\frac{r f'(r)+f(r)-1}{r^2}
&= -\frac{n a M^n}{r^{n+3}},
\label{ff1}
\\
\frac{f''(r)}{2}+\frac{f'(r)}{r}
&= \frac{n(n+1)aM^n}{2r^{n+3}}.
\label{ff2}
\end{align}
Considering the difference between Eqs.~(\ref{ff1}) and (\ref{ff2}), we obtain 
\begin{equation}
\dfrac{ f''(r)}{2}-\dfrac{ f(r)}{r^2}+\dfrac{1}{r^2}= \dfrac{n(n+3) \, a\, M^n}{2 r^{n+3}}. \label{ff22}
\end{equation}
This  resulting  equation  of  motion can  be solved as follows
\begin{equation}
f(r)=1+\dfrac{c_1}{r}+c_2 r^2+\dfrac{aM^n}{r^{n+1}}.
\end{equation}
To make contact with black holes solutions, we set 
$ c_1=-2M$ and  $c_2=-\dfrac{\Lambda}{3}.$
In this way,  the corresponding metric function takes the following form
\begin{equation}
f(r)=1-\frac{2M}{r}
\left(1-\frac{aM^{n-1}}{r^n}\right)
-\frac{\Lambda}{3}r^2.
\end{equation}
Considering  the limit $a=0$, we recover the ordinary Schwarzschild black hole  described with the  radial metric   function 
\begin{equation}
f(r)=1-\frac{2M}{r}
-\frac{\Lambda}{3}r^2.
\end{equation}
It should be noted that particular values of $n$  could recover  certain geometric solutions. They are, for instance, 
\begin{align}
n=& 1 \quad \longrightarrow \quad \text{NC black hole}\\
n=& 2 \quad \longrightarrow \quad \kappa\text{-deformation black hole}
\end{align}
studied in \cite{14,15}. Having constructed these black hole solutions, we move to investigate the associated thermodynamic behaviors. A close inspection shows that $n=1$ and $n=2$ are extremely studied in connection with NC and $\kappa-$deformed geometries relying on Lie algebraic structures, respectively \cite{14,15}. Here, however,  we consider models with $n>2.$ It has been observed that providing analytic expressions is a tough task. Nevertheless, it seems possible to approach the thermodynamic using parallel programming computation relaying on CUDA  code techniques. It  is recalled that  CUDA is a general purpose parallel computing platform   and programming model developped by  NVIDIA.  The GPU
architecture enables sharing the workloads between a number of streaming multiprocessors (SMs) in a highly parallel manner. Such  GPUs  provide  a multitude of powerful
methods and efficient tools to further exploit the architecture \cite{17,18}.   In a nutshell,   CUDA has proven to be a powerful tool in black hole physics to approach many properties including thermodynamics and optical shadows\cite{21,20}.

In what follows,  we consider three solutions corresponding to $n=3$, $n=4$, and $n=5$.  An examination shows that $n>5$ may require advanced numerical scenarios. This extension goes beyond the scope of the present work.

\begin{figure}[h!]
\centering
\begin{tabular}{c}
\includegraphics[width=1\linewidth]{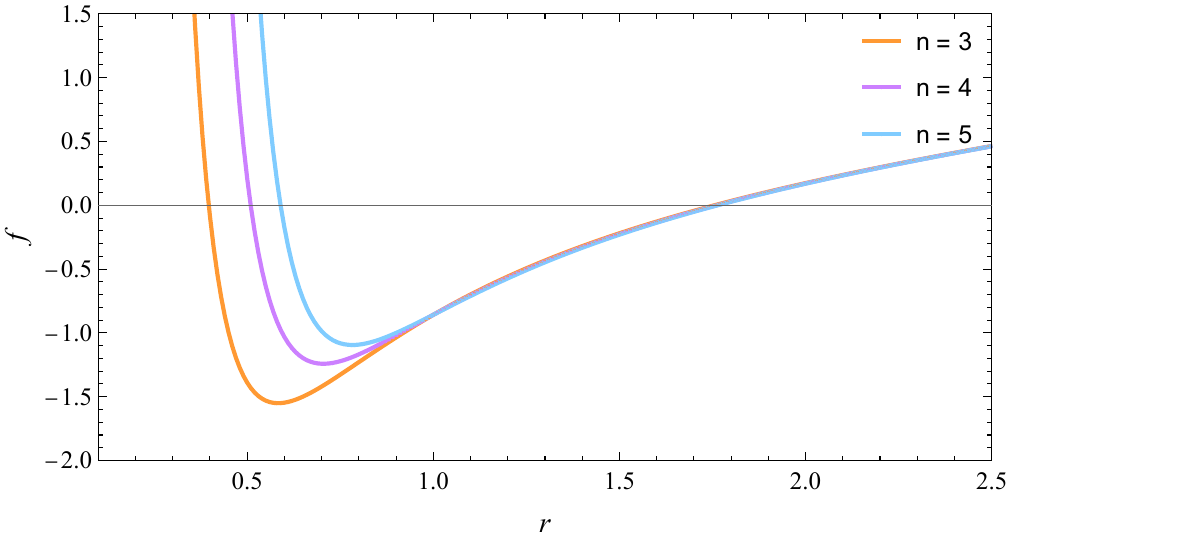} 
\end{tabular}
\caption{Metric function $f(r)$ behaviors.}
\label{f}
\end{figure}

Before approaching the thermodynamic  aspect,  we first  investigate the behavior of the metric function,  where certain parameters are fixed, thereby reducing the parameter space. We set the black hole mass to $M=1$, the deformation parameter to $a=0.1$, and the  cosmological constant to $\Lambda=-0.009$. Fig.~\ref{f} illustrates the metric function for three values of  $n$ being $n=3,4$ and $5$. The existence of double roots indicates the possibility of extremal black hole configurations.

\section{Numerical thermodynamic  compuations}
\begin{figure}[h]
\centering
\begin{tabular}{c}
\includegraphics[width=1\linewidth]{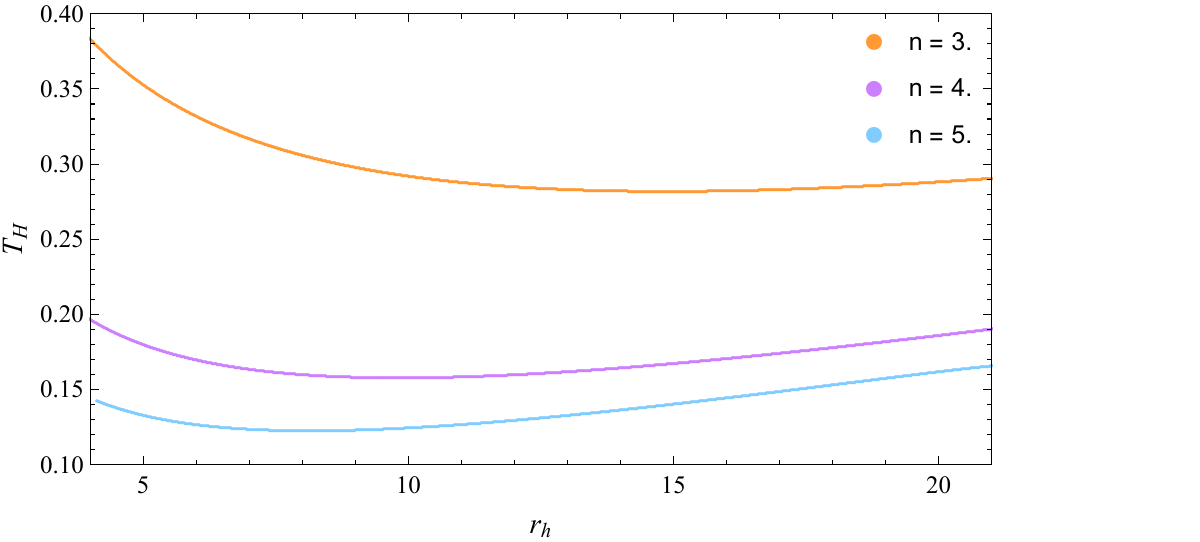}
\end{tabular}
\caption{ Hawking temperature $T_H$ as a function of the horizon radius $r_h$.}
\label{T}
\end{figure}

In this section,   we investigate stability behavior of such a  black hole solution. The first step is to determine its mass by imposing the horizon condition $f(r_h)=0$ given by
\begin{equation}
1-\frac{2M}{r_h}\left(1-\frac{aM^{n-1}}{2r_h^n}\right)
+\frac{8\pi P r_h^2}{3}=0
\end{equation}
where the cosmological constant has been identified with the pressure. Since this equation does not generally admit a simple analytical solution for arbitrary values of $n$, the mass  will be  determined numerically using CUDA parallel programming simulations. To this end, the parameters $a$ and $P$ are considered  fixed, while $n$ takes the values $3,4,$ and $5$. The numerical calculations are performed for different values of the horizon radius $r_h$. To do so, the Hawking temperature is expressed as follows
\begin{equation}
T_H=\left.\frac{1}{4\pi}\frac{\partial f(r)}{\partial r}\right|_{r=r_h}.\label{T1}
\end{equation}
To compute the Hawking temperature independently of the black hole mass, we first import the mass data obtained from the CUDA computations into the metric function, by  taking  $P=0.003$ and $a=-0.1$. Then, we  differentiate the metric function with respect to the horizon radius $r_h$. By substituting the resulting data into Eq. (\ref{T1}), we obtain the corresponding Hawking temperature values. In this regard, the Hawking temperature is plotted as a function of $r_h$ for different values of $n$, as illustrated in Fig. (\ref{T}).
 
This figure reveals that the Hawking temperature decreases to a minimum value for all three values of $n$. Then, it increases. An examination shows that the minimal value decreases by augmenting $n$. It has been observed that an increase in the parameter $n$ leads to lower values of the Hawking temperature.

\begin{figure}[h]
\centering
\begin{tabular}{c}
\includegraphics[width=1\linewidth]{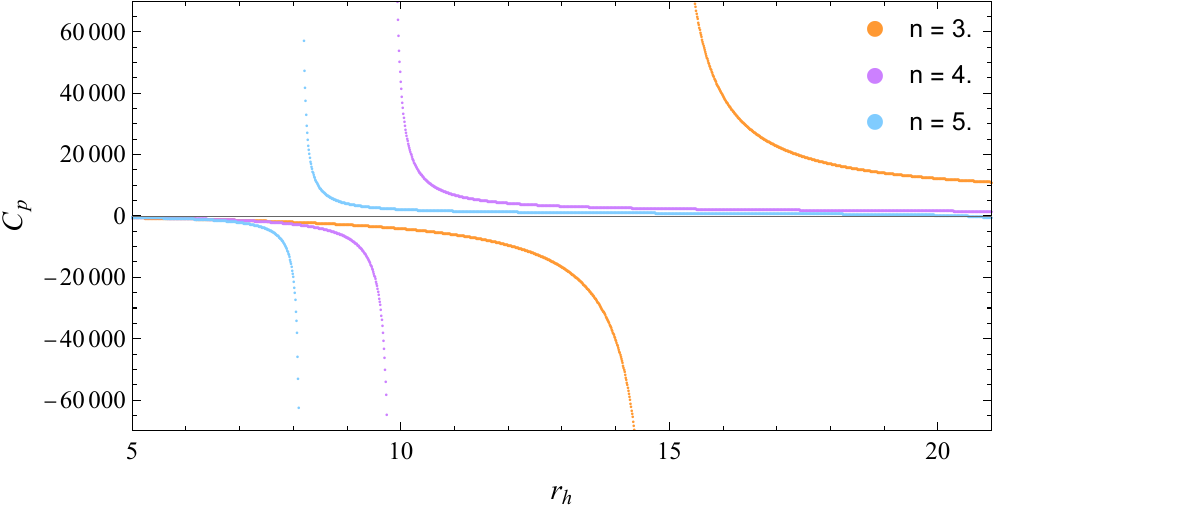}
\end{tabular}
\caption{ Heat capacity as a function of the horizon radius $r_h$.}
\label{cp}
\end{figure}

Having examined the thermal behavior of the black hole, we now turn to its thermodynamic stability. In particular, both local and global stability behaviors  are investigated by numerically   computing  the relevant thermodynamic quantities within the standard thermodynamic formalism. In order to approach the local stability, we need to compute the heat capacity via the
relation
\begin{equation}
C_{p}=\dfrac{\partial M}{\partial T_H}.\label{cp2}
\end{equation}
Using the output data files for the mass and the  Hawking temperature, we numerically compute their respective derivatives with respect to the horizon radius $r_h$ using parallel computing techniques with the help of CUDA codes. The resulting data are then substituted into Eq. (\ref{cp2}) in order to obtain the heat capacity.  In this way,  the heat capacity $C_p$ is plotted as a function of $r_h$ for different values of $n$, as shown in Fig. (\ref{cp})
 
\begin{figure}[h]
\centering
\begin{tabular}{c}
\includegraphics[width=1\linewidth]{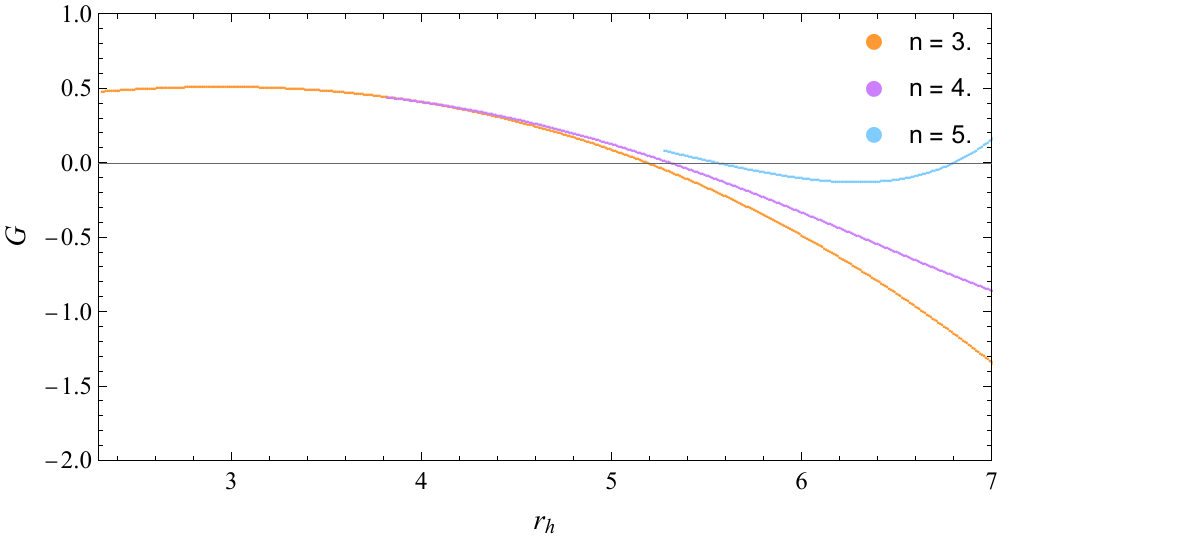}
\end{tabular}
\caption{ Gibbs free energy as a function of the horizon radius $r_h$.}
\label{G}
\end{figure}

It has been shown that the sign of the heat capacity can be used to control the stability of the corresponding black hole solutions. Indeed, a locally stable thermodynamic system corresponds to $C_p > 0$, whereas an unstable solution can occur if $C_p < 0$.
It has been observed from Fig.~(\ref{cp}) that the heat capacity curves exhibit discontinuous behaviors at certain critical horizon radii. Increasing the parameter $n$ leads to an increase in the critical values of $r_h$. At these points, the heat capacity exhibits divergent behaviors, clearly indicating a second-order phase transition.
After that, we investigate the global  stability by computing the Gibbs free energy, defined as
\begin{equation}
G=M-T_H S,\label{GG}
\end{equation}
where $S$ is the black hole entropy. The latter is  given by the Bekenstein–Hawking area law
\begin{equation}
S=\frac{A}{4}=\pi r_h^2.
\end{equation}
Here, $A$ is the area of the event horizon, which is expressed as $A=\left.\iint \sqrt{g_{\theta\theta}g_{\phi\phi}}\,d\theta d\phi\right|_{r=r_h}=4\pi r_h^2.$ 
Using this expression for the entropy, the mass,  and  the temperature data obtained from the CUDA code, we   numerically compute the Gibbs free energy. Fig.~\ref{G} shows the Gibbs free energy as a function of the horizon radius $r_h$ for different values of $n$, allowing  to examine the global thermodynamic stability of the black hole.

After discussing the stability behavior of the  obtain  black hole solution, we  move now to investigate their critical  and phase transition behaviors.
\section{$P-v$ criticality and phase transitions}
\begin{figure}[h]
\centering
\begin{tabular}{c}
\includegraphics[width=1\linewidth]{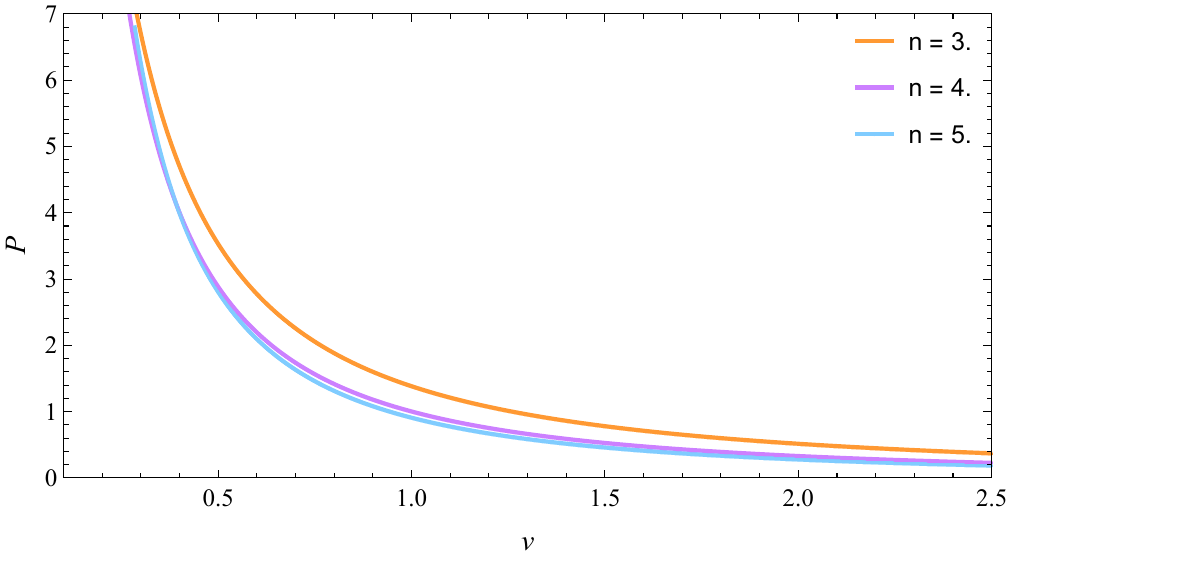}
\end{tabular}
\caption{ Pressure a function of the specific volume  $v$.}
\label{pv}
\end{figure}

In this section, we investigate the $P-v$ criticality of the black hole system to gain further insights into its thermodynamic properties and phase structure. In particular, we focus on identifying the critical points and examining the conditions under which the system exhibits phase transitions analogous to those of a Van der Waals fluid. This analysis provides a way to explore the influence of the model parameters on the critical behavior and the corresponding thermodynamic phase transitions. To  do so,  we derive the thermodynamic equation of state by combining the expressions for the Hawking temperature and  the thermodynamic pressure. Concisely,   the calculations  yield  the following equation of state 
\begin{equation}
P=\frac{3 \left(4\, \pi  T \,r_h^{2} -2 M  +a \,(1+n) M^{n} r_h^{-n}\right)}{16 \pi r_h^{3} }. \label{pp}
\end{equation}
In this way, the  critical points can determined from the equations
 \begin{equation}
\frac{\partial P}{\partial v }=0,\hspace{1.5cm}\frac{\partial ^{2}P}{%
\partial v^{2}}=0, \label{pv1}
\end{equation}
where $v$ represents the specific volume. To determine the critical pressure $P_c$, the  critical temperature $T_c$, and the  critical specific volume $v_c$, we employ parallel computing techniques to perform the needed  numerical calculations,  due  the complexity of the resulting  thermodynamic expressions. First, the mass and the  Hawking temperature data files are imported into Eq. (\ref{pp}) to obtain the corresponding pressure data. The resulting pressure data are then numerically differentiated with respect to $v$ to determine the critical points.  In this way, the corresponding values of $P_c$, $T_c$, and $v_c$ are obtained for different values of $n$. Fig. (\ref{pv}) illustrates the resulting $P-v$ criticality diagrams for different values of $n$ at $T=T_c$.

\begin{figure}[h]
\centering
\begin{tabular}{c}
\includegraphics[width=1\linewidth]{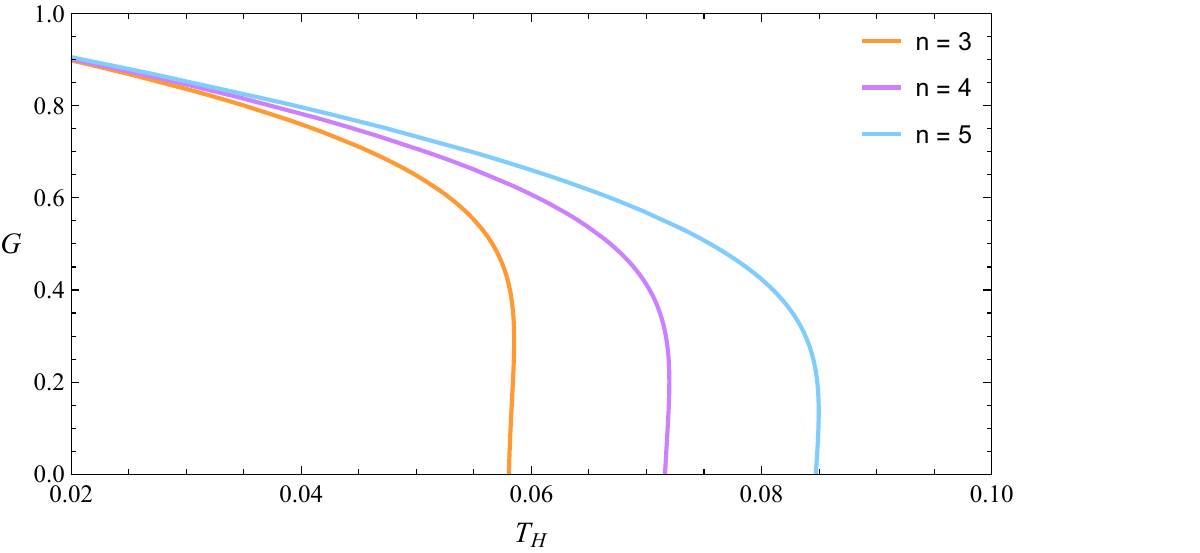}
\end{tabular}
\caption{ Gibbs free energy as a function of the Hawking temperature.}
\label{G1}
\end{figure}

Using the criticality results obtained above, we now turn to the investigation of the phase transitions of the black hole system. This analysis is essential for understanding the thermodynamic stability of the different black hole phases and identifying possible first-order phase transitions associated with the Van der Waals-like behavior. To this end, we consider the Gibbs free energy, given in Eq. (\ref{GG}), which serves as a useful tool for characterizing the thermodynamic phase structure. By fixing the pressure at its critical value $P=P_c$, we  numerically examine the variation of the Gibbs free energy as a function of the Hawking temperature. The corresponding behavior is illustrated in Fig.~(\ref{G1}).

It follows form this figure  that the Gibbs free energy decreases as the Hawking temperature increases. Increasing the parameter $n$, however,  shifts the curves towards higher temperatures. No swallowtail behaviors have been  observed, indicating the absence of a first-order phase transition in the present model.
\section{Conclusion}
Motivated by NC  geometry and related deformed geometrical frameworks, we have investigated a new class of asymptotically AdS black hole solutions characterized by  higher-order mass deformations  $\frac{aM^{n-1}}{r^n}$   governed by  two parameters   $a$ and $n$.  Due to  difficulties  associated with analytical   discussions,   we have  examined three different cases from a numerical perspective, namely $n=3$, $n=4$, and $n=5$, extending the investigation beyond the commonly studied cases $n=1$ and $n=2$. We have first  analyzed  the behavior of the obtained metric function and its dependence on the  deformation parameter.   Since obtaining thermodynamic  analytical  expressions for higher values of $n$ is particularly challenging, we have adopted a numerical approach supported by parallel computations using CUDA techniques. Precisely,   we have  numerically investigated the thermodynamic properties of  such extended  black holes by  focusing on the Hawking temperature, the entropy, and  the Gibbs free energy, as well as their stability and possible critical behaviors.  In a nutshell, we have shown  that the deformation parameters $n$ and $a$ have a significant influence on the thermodynamic characteristics of the system.  In particular, the Hawking temperature exhibits a minimum, whose value decreases with increasing $n$ and the deformation parameter $a$. Furthermore,  we have observed that the heat capacity reveals distinct regions of local thermodynamic stability and instability, with its divergences signaling the occurrence of second-order phase transitions. The Gibbs free energy  has been  also analyzed to investigate  the global thermodynamic stability and the possible competition between different black hole phases. In addition, the $P-v$ criticality analysis, combined with the behavior of the Gibbs free energy at the critical pressure has provided  further insight into the phase structure and the possible emergence of Van der Waals-like phase transitions of such models. 

In the end of this work, we would  like to provide a concise speculation  on the involved parameter $a$  appearing in  the mass  correction term.  It has been remarked the interpretation of  such a parameter depends of the possible  values of    $n$.  For $n=1$ and $  n=2$,  for instance,  it has been linked to the constant and linear spacetime algebra  structures,  respectively. Supported by such a geometric aspect,  we could intuitively anticipate a possible interpretation of higher-degree commutator bracket algebras   of   the form  $$ [x^\mu, x^\nu]= \sum\limits_{i+j+k+\ell=n-1}  a_{ijk\ell} C^{\mu\nu}_{\sigma\rho \delta \gamma} (x^{\sigma})^i (x^\rho)^j (x^\delta)^k ( x^\gamma)^\ell$$ where one has used  $i,j,k,\ell=0,1$ and the remaining ones are spacetime indices.  This suggested intuition  may require a pensive  consideration on the involved tensor forms, which should be left for future investigations.

{\bf Data availability}\\
  The data are available from the corresponding author upon reasonable request. \\

{\bf Acknowledgements}\\ 
MJ  gratefully acknowledges the financial support of the CNRST in the
frame of the PhD Associate Scholarship Program PASS.


\begin{thebibliography}{10}

\bibitem{1} K. Schwarzschild, \textit{On the Gravitational Field of a Mass Point According to Einstein's Theory},  Sitzungsber. Preuss. Akad. Wiss. Berlin (Math.Phys.) 189 (1916), \texttt{arXiv:physics/9905030}.


\bibitem{2} J. D. Bekenstein,  \textit{Black Holes and Entropy}, Physical Review D, 7, 2333 (1973).

\bibitem{3} S. W. Hawking, \textit{Black Holes and Thermodynamics}, Physical Review D, 13, 191 (1976), \texttt{	arXiv:gr-qc/9702022}. 

\bibitem{4} R. M. Wald, \textit{The Thermodynamics of Black Holes}, Living Rev.Rel. 4, 6 (2001), \texttt{arXiv:gr-qc/9912119}.

\bibitem{5} S. Sachdev  and M. Müller, \textit{Quantum Criticality and Black Holes},  J. Phys. Condens. Matter 21, 164216 (2009), \texttt{arXiv:0810.3005 [cond-mat.str-el]}.

\bibitem{6} K. Bradler, C. Adami, \textit{The capacity of black holes to transmit quantum information}, JHEP 05,095 (2014), \texttt{	arXiv:1310.7914 [quant-ph].}

\bibitem{7} M. Franz and M. Rozali, \textit{Mimicking Black Hole Event Horizons in Atomic and Solid-State Systems},  Nature Rev.Mater. 3, 491 (2018), \texttt{arXiv:1808.00541 [cond-mat.str-el]}.
\bibitem{07} C. Sims,  \textit{Analogous Black Holes in Type-III Dirac Semimetal Ni3In2X2 (X = S, Se)},  \texttt{
arXiv:2304.00642}.
\bibitem{70} 
J. D. Bekenstein, \textit{Extraction of Energy and Charge from a Black Hole, }Phys. Rev. D7, 949 (1973).
\bibitem{8} S. W. Hawking, D. N. Page, \textit{Thermodynamics of Black Holes in anti-de Sitter Space}, Commun. Math.  Phys. 87, 577 (1983).

\bibitem{9} A. Belhaj, M. Jemri, \textit{On Thermodynamics of Charged Black Holes via Extended Space-time Derivatives}, Int.J.Mod.Phys.A 41, 04, 2650027 (2026), \texttt{	arXiv:2511.18407 [hep-th]}.

\bibitem{10}   M.  Jemri,  \textit{Dunkl-Corrected Deformation of RN-AdS Black Hole Thermodynamics}, Theor.Math.Phys. 22, 1, 720   (2026),   \texttt{	arXiv:2512.19200}. 

\bibitem{11} D. Kubiznak, R. B. Mann, \textit{P-V Criticality of Charged AdS Black Holes}, Physical Review D, 84, 024037 (2011), \texttt{arXiv:1106.5789}.
\bibitem{110}
A. Belhaj, M. Chabab, H. El Moumni, M. B. Sedra,   \textit{On Thermodynamics of AdS Black Holes in Arbitrary Dimensions}, Chin.Phys.Lett. 29 100401(2012),   \texttt{arXiv:1210.4617}.

\bibitem{12} A. Belhaj, M. Jemri, \textit{On Thermodynamic Universalities and Optics of B-deformed Reissner--Nordström--AdS Black Holes}, \texttt{	arXiv:2609.25161 [hep-th]}.

\bibitem{13} H. Belmahi, M. Jemri, R. Salih, \textit{Stability and Criticality Behaviors of Accelerating Charged AdS Black Holes in Rainbow Gravity},  Int. J. Mod. Phys. A 40,35, 2550149 (2025), \texttt{	arXiv:2507.03572 [gr-qc]}.


\bibitem{014} 
 A. A. Araújo Filho, J. R. Nascimento, A. Yu. Petrov, P. J. Porfírio, Ali Övgün, \textit{Effects of non-commutative geometry on black hole properties},  Phys.Dark Univ. 46 101630 (2024), \texttt{arXiv:2406.12015}. 
 \bibitem{0114} 
 F. Rahaman, A. Banerjee, I. Radinschi, M. Kalam, S. Islam, \textit{Noncommutative geometry
inspired charged black holes}, Physics Letters B, 694(1) 10 (2010).
\bibitem{14} W. El Hadri, M. Jemri, \textit{Thermodynamics and Criticality of Noncommutative RN-AdS Black Holes}, Braz J Phys 56, 23 (2026), \texttt{		arXiv:2509.00926 [hep-th]}.

\bibitem{15} A. Naveena Kumara, V. Rajagopal and P. Wu, \textit{Thermodynamics and Phase Transitions of $\kappa$-Deformed Schwarzschild-AdS Black Holes}, Physics Letters B, 880, 140724 (2026), \texttt{	arXiv:2604.08629 [gr-qc]}.


\bibitem{015}
B. E. Panah,   $\kappa$-deformed Schwarzschild AdS black strings,   \texttt{arXiv:2609.30994}.
\bibitem{16} V. V. Kiselev, \textit{Quintessence and black holes}, Class.Quant.Grav. 20, 1187 (2003), \texttt{		arXiv:gr-qc/0210040}.

\bibitem{17} G. Quiroga,\textit{ MALBEC: a new CUDA-C ray-tracer in general relativity}, (2018)
\bibitem{18} A. G. M. Lewis, H. P. Pfeiffer, \textit{GPU-Accelerated Simulations of Isolated Black Holes},
Class. Quant. Grav. 35 (2018) 095017, 
\texttt{arXiv:1804.09101 [gr-qc].}
\bibitem{21} A. Belhaj,  M. Jemri, \textit{On thermodynamic universals of black holes in machine learning with CUDA}, Astronomy and Computing,101210(2026).
\bibitem{20} S. E. Baddis, A. Belhaj, H. Belmahi, M. Jemri, \textit{Constraining Black Hole Shadows in Dunkl Spacetime using CUDA Numerical Computations},
JHEAp 51, 100541 (2026), 
\texttt{arXiv:2510.16460 [gr-qc].}

\end{thebibliography}
\end{document}